# Self-organized compound pattern and pulsation of dissipative solitons in a passively mode-locked fiber laser

Zhenhong Wang[1], Zhi Wang[1*], Yange Liu[1], Ruijing He[1], Jian Zhao[2], Guangdou Wang[1], Shangcheng Wang[1], Guang Yang[1]

[1]Key Laboratory of Optical Information and Technology, Ministry of Education, Tianjin Key Laboratory of Optoelectronic Sensor and Sensing Network Technology, and Institute of Modern Optics, Nankai University, Tianjin, 300350, China

[2]Key Laboratory of Opto-electronic Information Technical Science of Ministry of Education, School of Precision Instruments and Optoelectronics Engineering, Tianjin University, Tianjin, 300072, China

[*]zhiwang@nankai.edu.cn

**Abstract:** We experimentally observe soliton self-organization and pulsation in a passively mode-locked fiber laser. The optomechanical interaction in the optical fiber is key to the formation of equidistant soliton bunches. These solitons simultaneously undergo a pulsation process with a period corresponding to tens of the cavity round trip time. Using the dispersive Fourier transformation technique, we find that the Kelly sidebands in the shot-to-shot spectra appear periodically, synchronizing with the pulsation.

## 1. Introduction

The evolution and interaction of a large number of solitons dominate in various nonlinear dynamic processes, such as those of Bose-Einstein condensates [1], plasmas [2], optical systems [3, 4], and complex network [5]. Because of the energy quantization effect of a dissipative system, numerous solitons tend to emerge in passively mode-locked fiber lasers (PMLFLs) with an anomalous dispersion cavity [6]. These solitons can self-organize into stable or unstable sequences on a global and local scale via various interaction mechanisms. PMLFLs with multiple soliton patterns constitue excellent ultrafast pulse sources for data storage [7], micromachining and imaging [8], and provide an ideal platform for the fundamental exploration of intricate dissipative nonlinear dynamics [4]; thus, they have been investigated extensively.

Self-organizing soliton processes are usually dominated by different interaction mechanisms. Direct interaction through coherent overlapping of the pulse tail can form a stable bound state, such as a soliton crystal [9], whereas the dispersive wave emitted by a soliton usually contributes to the formation of a loose soliton bunch in the medium range [10]. On the other hand, long-range interactions, including the optomechanical effect [11, 12], the slow depletion and recovery processes of the gain [13], the quasi-continuous-wave (cw) component [14, 15], and the noise floor [16], can induce soliton redistribution throughout the cavity. This behavior may induce the formation of global structures, such as harmonic mode-locking (HML) [7] and soliton rain [17]. Moreover, when a large number of solitons emerge in PMLFLs, composite patterns in both global and local ranges form through the involvement of multiple interaction mechanisms [18, 19]. Thus, the generation of composite patterns involves manifold interaction mechanisms. However, the details of the internal motions of multiple solitons have rarely been studied; thus, further analysis is required, particularly using fast real-time measurement methods. Further, because of the parameter space complexity and the time window limitation, numerical investigation of these composite patterns remains challenging. Therefore, systemic experimental work is always very important for direct revelation of the global dynamics of large numbers of solitons in PMLFLs.

Soliton pulsation, which is related to a periodic attractor in a nonlinear dissipative system, is usually regarded as an intermediate regime between stable solitons and chaos. Depending on the evolutionary processes, various pulsation behaviors, such as plain, erupting, creeping, and so-called "strong" pulsation have presented in the parameter space of the cubic-quintic complex Ginzburg-Landau equation [20 22]. Pulsation phenomena have also been observed experimentally in response to soliton energy modulation [21]. In contrast, the evolutionary behavior of solitons during the pulsation process has rarely been investigated experimentally.

Using conventional detection methods, it is difficult to gain a direct insight into the internal characteristics and pulsation of soliton bunches. Capturing these bunches requires real-time spectral and temporal measurements. The dispersive Fourier transformation (DFT) technique can overcome the speed limitation of a conventional optical spectrum analyzer (OSA) and facilitate fast real-time measurements [23]; thus, DFT has been widely used in fiber lasers for effective resolution of the evolution of unstable processes, including noise-like pulses [24], unstable soliton bunches [25], and rogue waves [26].

In this letter, we report a remarkable experimental observation of soliton self-organization and pulsation in a PMLFL with a graphene saturable absorber (GSA) on a microfiber. By tuning the polarization state in the PMLFL, multiple solitons separate from a condensed phase and spontaneously form soliton bunches, which are periodically distributed throughout the cavity by the optomechanical interaction. Using the DFT technique, the internal characteristics of the soliton bunches are investigated. The solitons have fixed phase relations but varying separations, and the average number per bunch increases with the pump power. These solitons pulsate at a period of tens of the cavity round trip time. Kelly sidebands appear periodically in the shot-to-shot spectra, accompanying the pulsation.

**2. Experimental setup**

The experimental setup of the PMLFL with the fabricated microfiber-based GSA is shown in Fig. 1. Note that the microfiber-based GSA was constructed via the deposition method [27]. A 2.8-m-long erbium-doped fiber (EDF) was used as the gain medium, pumped by a 980-nm laser diode (LD) through a fused wavelength division multiplexer (WDM). A polarization-independent isolator (PI-ISO) maintained the unidirectional laser pulse propagation and a polarization controller (PC) was used to tune the cavity polarization state. The laser output was directed through the 10% port of an optical coupler (OC). The total cavity length was ~16.4 m. The output pulse trains and operational real-time performance were evaluated by a 50-GHz bandwidth photodetector (U2T XPDV2120R) and a 36-GHz oscilloscope with an 80-GS/s maximal sampling rate (LeCroy LCRY3312N69578).

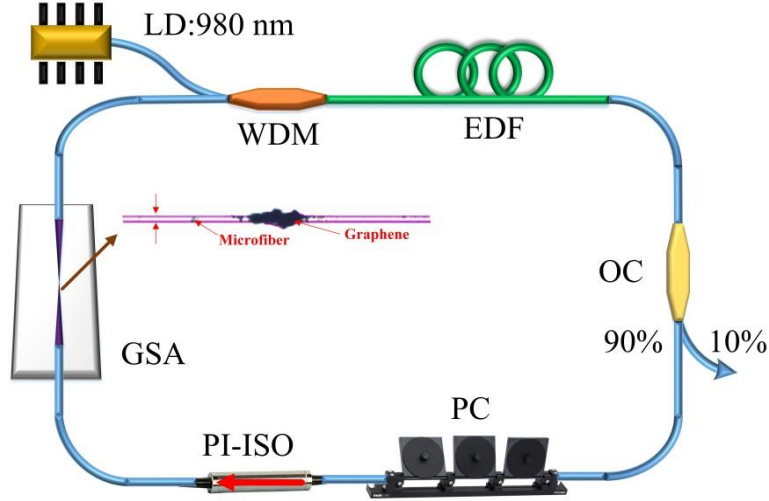

Fig. 1. Schematic setup of PMFL with a graphene saturable absorber (GSA) on microfiber.

**3. Experimental results and discussion**

For a pump power of more than 108 mW, we can observe the conventional mode-locked soliton by adjusting the PC in the laser cavity appropriately. Further, by adjusting one PC paddle, the conventional soliton regime can be translated to a reverse soliton rain [17]. In this study, the pulse-train evolutionary process (Fig. 2) was recorded using a 2-GHz oscilloscope (LeCroy WaveRunner 620Zi) at 160-mW pump power. As shown in Fig. 2, solitons arose spontaneously from the condensed phase, drifted right, and gradually disappeared. This phenomenon can be regarded as evaporation of the condensed soliton phase into soliton gas. When we further adjusted the PC paddle, the condensed phase faded and the pulses were eventually frozen at almost equal intervals throughout the entire cavity.

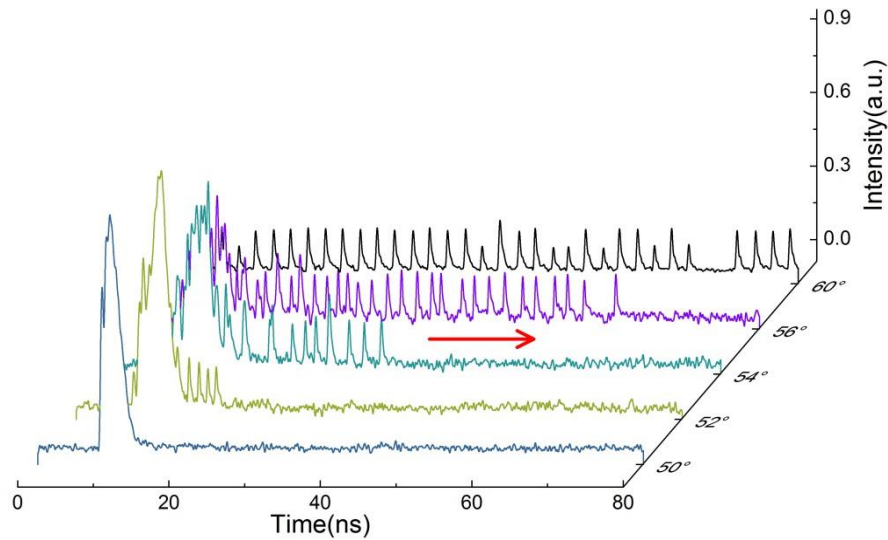

Fig. 2. Soliton pattern evolution in cavity round with PC angle change.

We further observed the final state using a high-speed oscilloscope (36 GHz, 80 GS/s). Figs. 3(a)-3(c) show the three-dimensional spatiotemporal diagrams of an 82-ns single cavity round trip time (~9762 cavity cycles) at pump powers of 108, 160, and 360 mW, respectively. Magnifying these figures on the time scale, we found that each wave packet was composed of one or more solitons (soliton bunch; Figs. 3(d)-3(f)). Note that the bunches with different soliton numbers appeared as pulses with different peak powers in the low-speed oscilloscope (Fig. 2). Most soliton bunches had the same spacing (~2.28 ns) and were independent of the pump power. Note that this phenomenon is very similar to HML, except some wider intervals appear in the cavity.

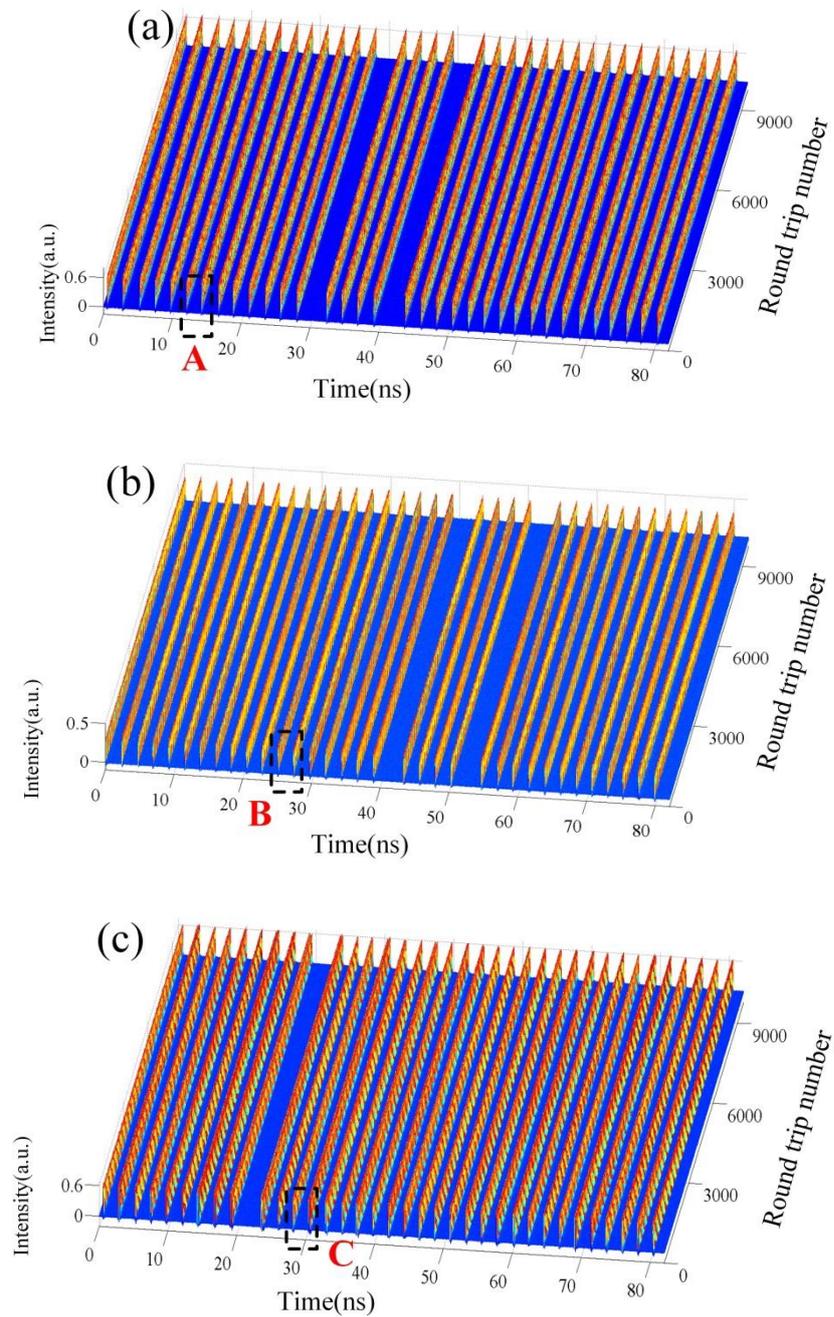

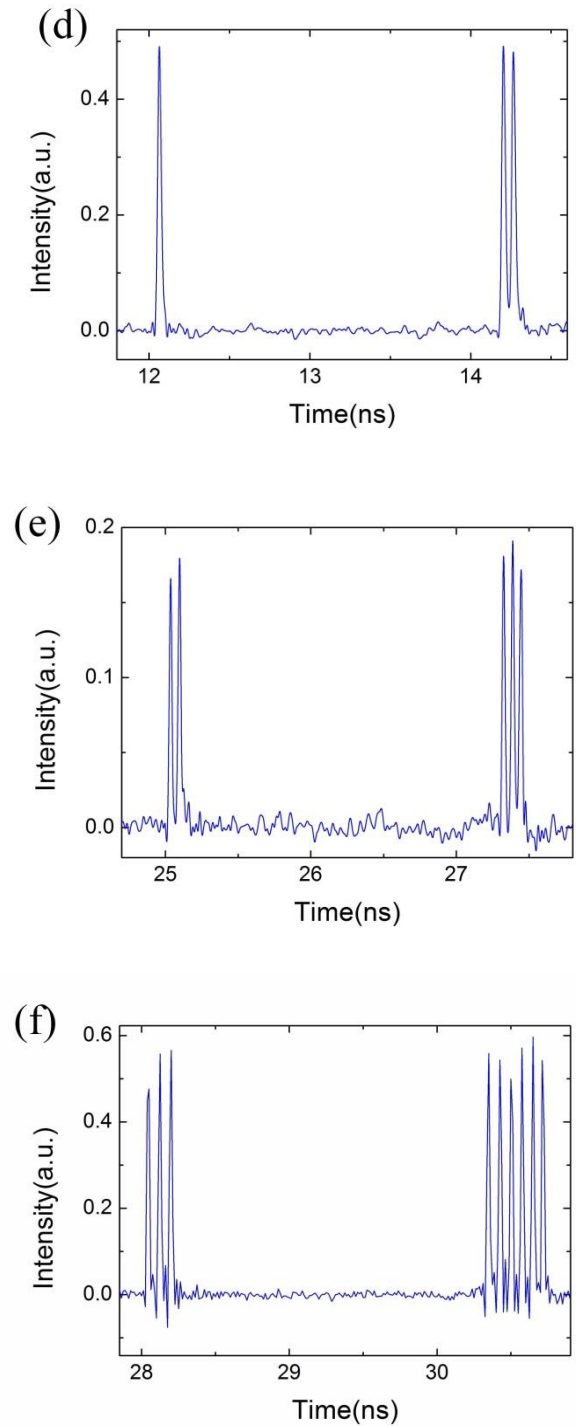

Fig. 3. Spatiotemporal diagrams for pump powers of (a) 108, (b) 160, and (c) 360 mW. (d)-(f) Temporal pulse trains in sections A-C for first cavity round trip, respectively.

The soliton self-organization throughout the cavity is attributed to the optomechanical interaction of soliton bunches in the optical fibers. Solitons typically generate an intense electric field that distorts density changes in the fiber and excites acoustic waves that propagate transverse to the fiber axis [12]. The optical fiber then acts as a resonator for the acoustic waves excited at the appropriate natural eigenfrequencies of the fiber. When the soliton sequence has the same frequency as a specific acoustic eigenmode, the resonance enhances the acoustic wave intensity and locks the spacing between the solitons. In this experiment, the soliton bunch frequency was ~439 MHz (2.28 ns), essentially agreeing with the eigenfrequencies of the optical-fiber acoustic mode [12]. The wider intervals were ~4.38 ns, slightly less than double 2.28 ns. Therefore, we speculate that these intervals were due to the slight frequency mismatch between the laser cavity period and the acoustic-mode eigenfrequency.

To confirm this hypothesis, we appended an ~4-cm single mode fiber (SMF) onto the cavity. Then, the soliton bunches underwent almost the same self-organizing processes, but the wider intervals disappeared and conventional HML was achieved. When the cavity length was increased further, a stable periodic pattern of soliton bunches could not be realized until the resonance with the other discrete acoustic eigenfrequency was satisfied. Note that microfiber has a similar structure to the fiber core of a silica-air photonic crystal fiber (so as to enhance the acoustic wave); thus, the impact of the optomechanical effect on the microfiber was also investigated by fabricating a GSA with microfiber diameters of 6, 8, and 10 m. When the GSA with different diameters was substituted into the apparatus and the cavity length was fixed, the soliton bunch frequency was invariable; thus, the microfiber was too short to impact the soliton bunch distribution.

Besides the optomechanical interaction, the gain depletion and recovery [13] and the quasi-cw component [15] have also been used to explain the HML form in numerous theoretical and experimental studies. The long-range repulsive interaction generated by the slow gain depletion and recovery effect can drive the solitons to form an equidistant distribution throughout the cavity. This effect has been regarded as a dominant factor in HML formation in many PMLFL configurations. Here, however, the uniform inter-soliton repulsive interaction fails to explain the existence of the wider intervals among the uniform distribution of the soliton bunches, as well as the difference in the soliton numbers of different bunches. Moreover, the soliton bunch spacing is independent of the pump power. These phenomena indicate that gain depletion and recovery is not a major factor in the self-organization of the soliton bunches in the PMLFL. Further, as shown in Fig. 4(a), when the soliton bunches stopped moving in the cavity, no quasi-cw component was apparent in the spectra. Hence, we believe that the optomechanical interaction in the optical fiber has a dominant effect on the formation of the global distribution of the soliton bunches in the PMLFL.

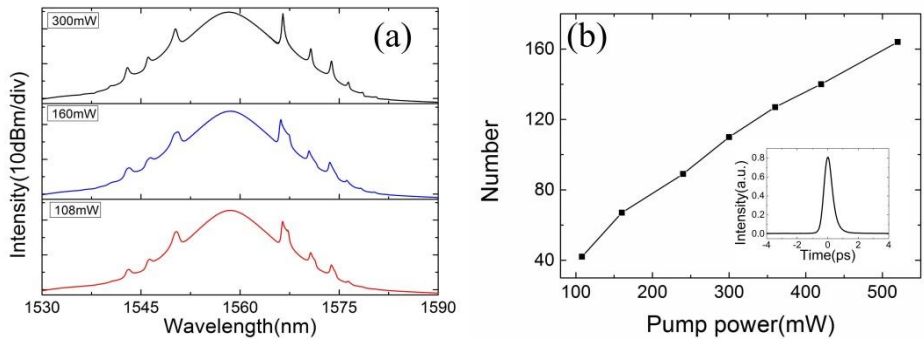

Fig. 4 (a) OSA-detected optical spectra at different pump powers. (b) Soliton numbers in one cavity round trip with different pump powers. (b, inset) Measured autocorrelation at 160-mW pump power.

Next, we focused on the internal soliton bunches. Each soliton bunch contained one or several solitons, as shown in Figs. 3(d)-3(f). Further, the soliton number in one cavity round trip gradually increased with increasing pump power (Fig. 4(b)), as a result of the soliton energy quantization effect. As the bunch number was almost fixed, the average soliton number in each bunch also increased with increasing pump power. Moreover, the pulse separations in the soliton bunches were inequivalent, varying from 63 to 77 ps. Fig. 4(b, inset) shows the autocorrelation trace at 160-mW pump power, which indicates that the soliton width was ~640 fs. As the soliton spacing was significantly larger than the width, the direct pulse-to-pulse interactions were very weak and the soliton bunches were loose. Fig. 4(a) shows the optical spectra of the OSA at different pump powers. These spectra are smooth with obvious Kelly sidebands, very similar to those of traditional soliton mode-locking.

To further investigate the detailed characteristics of the soliton bunches, we measured the shot-to-shot optical spectra of 9762 cavity cycles using the DFT technique. The soliton bunches (consisting of many solitons) were too wide for the temporal information to be transformed to a spectrum via the DFT technique; thus, we selected a single soliton or dual-soliton bunch only to analyze the shot-to-shot spectra. For bunches having more solitons, the DFT technique cannot directly obtain the spectral information; however, very detailed information on the soliton bunches can be acquired [28-30].

Figs. 5(a) and 5(b) show the evolutions of the optical spectra over 400 consecutive roundtrips of a single soliton and a dual-soliton bunch, corresponding to Figs. 3(d, e; left), respectively. Distinct periodic changes, or soliton pulsations, are apparent in the spectra. To more clearly indicate the pulsation process, the single-shot spectra at the roundtrips with maximum and minimum intensities in a pulsation period are shown in Figs. 5(c) and 5(d). The averaged spectra over 9762 cavity cycles are also illustrated in these figures, being basically consistent with the OSA spectra (Fig. 4(a)). To characterize the pulsation process, we calculated the intracavitary energy Q by integrating the shot-to-shot spectral power in each cavity period. Fig. 6(a) shows that Q varies almost sinusoidally with ~87 and ~80 cavity periods at 108- and 160-mW pump powers, respectively. These pulsation periods match the frequency difference between the main line and satellites in the radio frequency (RF) spectra shown in Figs. 6(c) and 6(d). In addition, the pulsation periods gradually decline with increasing pump power (Fig. 6(b)).

A notable phenomenon in the pulsation process is the periodic appearance of Kelly sidebands in the shot-to-shot spectra. The Kelly sidebands correspond to the dispersive waves in the temporal domain, which, in the usual perspective, radiate from solitons when they are perturbed by lumped nonlinear losses and various intracavity components in a round trip [10]. Here, the dispersive wave experiences a periodic radiation during tens of cavity periods. The energy of the radiated dispersive wave corresponds to a significant proportion of the change in pulsation energy. To the best of our knowledge, this pulsation phenomenon has not yet been reported in theory or experiment. We speculate that the pulsation may be related to the combination of the lumped nonlinear loss of the GSA and the slow gain depletion and recovery process. However, further theoretical investigation should be conducted in order to reveal its physical mechanism.

As illustrated in Figs. 5(b) and 5(d), the shot-to-shot spectra of the dual-soliton bunch exhibit clear interference fringes. These fringes can also be found in bunches with greater numbers of solitons, indicating that multiple solitons in the same bunch have fixed phase differences. Note that, because of the large pulse spacing, the interference fringes are not apparent in the OSA spectra shown in Fig. 4(a). The large but unequal spacing, as well as the fixed phase relation, implies that the periodically radiated dispersive wave could play a crucial role in the soliton bunch formation [10].

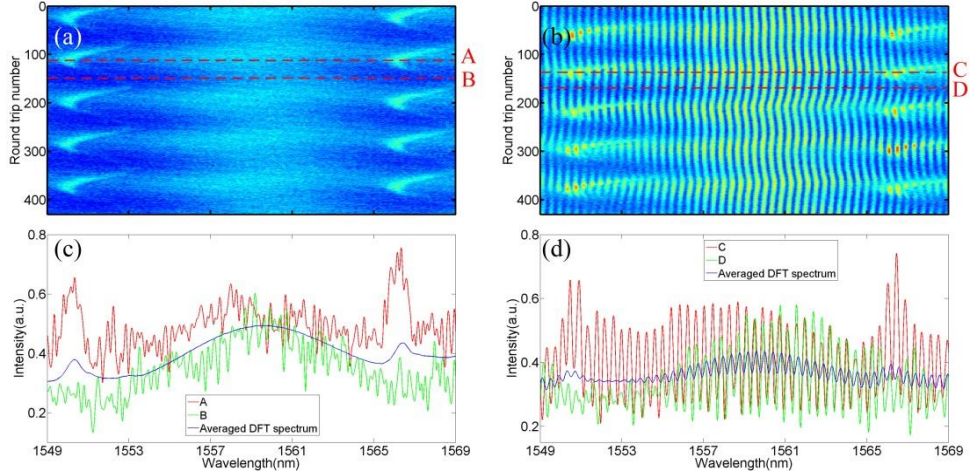

Fig. 5. Consecutive shot-to-shot spectral sequences for (a) one soliton at 108 mW [corresponding to left soliton in Fig. 3(d)], and (b) dual-soliton bunch at 160 mW [corresponding to left soliton bunch in Fig. 3(e)]. (c) Single-shot spectra at round trips A (red) and B (green) in (a). (d) Single-shot spectra at round trips C (red) and D (green) in (b). The averages (blue) of 9762 consecutive single-shot spectra in both cases are also shown in (c) and (d), respectively.

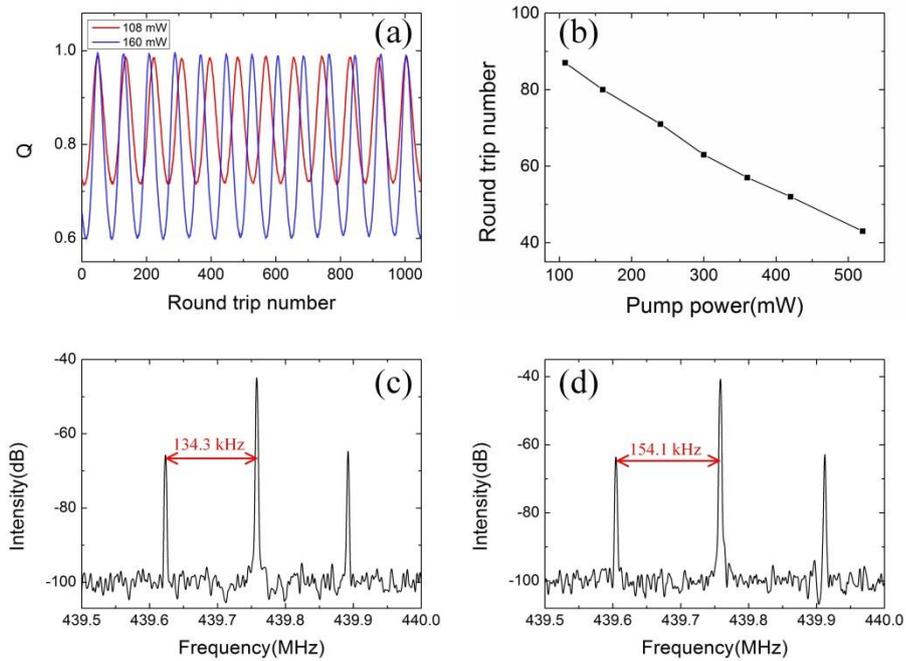

Fig. 6. (a) Periodic evolution of intracavitary energy Q at 108- and 160-mW pump powers, respectively. (b) Round trip numbers of soliton pulsations at different pump powers. (c, d) Measured RF spectra at 108- and 160-mW pump powers, respectively.

## 4. Conclusion

In summary, we report a remarkable experimental observation of soliton self-organization and pulsation in PMLFL with a GSA on a microfiber. According to our analysis, the optomechanical interaction, which is usually considered very weak in conventional optical fibers, is key to the formation of equidistant soliton bunches via self-organized processes of a large number of solitons in the PMLFL. Moreover, via the DFT technique, we investigated the internal characteristics and pulsation of the soliton bunches. Hence, we demonstrated that Kelly sidebands appear periodically in the shot-to-shot spectra, synchronizing with the pulsation process; thus, the dispersive wave experiences periodic radiation during tens of cavity periods. This pulsation process has rarely been studied in theory or experiment. Further explanation of the formation of the observed soliton pulsations will be the aim of our future work. We expect that our results can promote further theoretical and experimental research to explore the undiscovered mechanisms behind the distribution and evolution of large numbers of solitons in dissipative systems.

**Acknowledgments**

This work was supported by the National Natural Science Foundation of China (NSFC) (11674177, 61322510, 61640408) and Tianjin Natural Science Foundation (16JCZDJC31000, 14JCZDJC31300).

**References**

1. J. Denschlag, J. Simsarian, D. Feder, C. W. Clark, L. Collins, J. Cubizolles, L. Deng, E. W. Hagley, K. Helmerson, and W. P. Reinhardt, "Generating solitons by phase engineering of a Bose-Einstein condensate," Science **287**, 97-101 (2000).

2. N. J. Zabusky, and M. D. Kruskal, "Interaction of" solitons" in a collisionless plasma and the recurrence of initial states," Physical review letters **15**, 240 (1965).

3. J. M. Dudley, F. Dias, M. Erkintalo, and G. Genty, "Instabilities, breathers and rogue waves in optics," Nature Photonics **8**, 755-764 (2014).

4. P. Grelu, and N. Akhmediev, "Dissipative solitons for mode-locked lasers," Nature Photonics **6**, 84-92 (2012).

5. S. H. Strogatz, "Exploring complex networks," Nature **410**, 268-276 (2001).

6. A. Grudinin, D. Richardson, and D. Payne, "Energy quantisation in figure eight fibre laser," Electronics Letters **28**, 67-68 (1992).

7. M. Pang, W. He, X. Jiang, and P. S. J. Russell, "All-optical bit storage in a fibre laser by optomechanically bound states of solitons," Nature Photonics (2016).

8. M. E. Fermann, and I. Hartl, "Ultrafast fibre lasers," Nature photonics **7**, 868-874 (2013).

9. F. Amrani, A. Haboucha, M. Salhi, H. Leblond, A. Komarov, and F. Sanchez, "Dissipative solitons compounds in a fiber laser. Analogy with the states of the matter," Applied Physics B: Lasers and Optics **99**, 107-114 (2010).

10. A. Komarov, F. Amrani, A. Dmitriev, K. Komarov, D. Meshcheriakov, and F. Sanchez, "Dispersive-wave mechanism of interaction between ultrashort pulses in passive mode-locked fiber lasers," Physical Review A **85**, 013802 (2012).

11. A. Pilipetskii, E. Golovchenko, and C. Menyuk, "Acoustic effect in passively mode-locked fiber ring lasers," Optics letters **20**, 907-909 (1995).

12. A. B. Grudinin, and S. Gray, "Passive harmonic mode locking in soliton fiber lasers," Journal of the Optical Society of America B **14**, 144-154 (1997).

13. J. N. Kutz, B. Collings, K. Bergman, and W. Knox, "Stabilized pulse spacing in soliton lasers due to gain depletion and recovery," IEEE journal of quantum electronics **34**, 1749-1757 (1998).

14. D. Tang, B. Zhao, L. Zhao, and H. Tam, "Soliton interaction in a fiber ring laser," Physical Review E **72**, 016616 (2005).

15. Z. X. Zhang, L. Zhan, X. X. Yang, S. Y. Luo, and Y. X. Xia, "Passive harmonically mode-locked erbium-doped fiber laser with scalable repetition rate up to 1.2 GHz," Laser Physics Letters **4**, 592–596 (2007).

16. R. Weill, A. Bekker, V. Smulakovsky, B. Fischer, and O. Gat, "Noise-mediated Casimir-like pulse interaction mechanism in lasers," Optica **3**, 189-192 (2016).

17. S. Chouli, and P. Grelu, "Soliton rains in a fiber laser: An experimental study," Physical Review A **81**, 063829 (2010).

18. F. Amrani, A. Niang, M. Salhi, A. Komarov, H. Leblond, and F. Sanchez, "Passive harmonic mode locking of soliton crystals," Optics letters **36**, 4239-4241 (2011).

19. Y.-Q. Huang, Z.-A. Hu, H. Cui, Z.-C. Luo, A.-P. Luo, and W.-C. Xu, "Coexistence of harmonic soliton molecules and rectangular noise-like pulses in a figure-eight fiber laser," Optics Letters **41**, 4056-4059 (2016).

20. J. M. Soto-Crespo, N. Akhmediev, and A. Ankiewicz, "Pulsating, creeping, and erupting solitons in dissipative systems," Physical review letters **85**, 2937 (2000).

21. J. M. Soto-Crespo, M. Grapinet, P. Grelu, and N. Akhmediev, "Bifurcations and multiple-period soliton pulsations in a passively mode-locked fiber laser," Physical Review E **70**, 066612 (2004).

22. W. Chang, J. M. Soto-Crespo, P. Vouzas, and N. Akhmediev, "Extreme soliton pulsations in dissipative systems," Physical Review E **92**, 022926 (2015).

23. K. Goda, and B. Jalali, "Dispersive Fourier transformation for fast continuous single-shot measurements," Nature Photonics **7**, 102-112 (2013).

24. C. Lecaplain, and P. Grelu, "Rogue waves among noiselike-pulse laser emission: an experimental investigation," Physical Review A **90**, 013805 (2014).

25. Z. Wang, Z. Wang, Y.-g. Liu, W. Zhao, H. Zhang, S. Wang, G. Yang, and R. He, "Q-switched-like soliton bunches and noise-like pulses generation in a partially mode-locked fiber laser," Optics Express **24**, 14709-14716 (2016).

26. Z.-R. Cai, M. Liu, S. Hu, J. Yao, A.-P. Luo, Z.-C. Luo, and W.-C. Xu, "Graphene-Decorated Microfiber Photonic Device for Generation of Rogue Waves in a Fiber Laser," IEEE Journal of Selected Topics in Quantum Electronics **23**, 1-6 (2017).

27. K. Kashiwagi, and S. Yamashita, "Deposition of carbon nanotubes around microfiber via evanescent light," Optics express **17**, 18364-18370 (2009).

28. G. Herink, F. Kurtz, B. Jalali, D. Solli, and C. Ropers, "Real-time spectral interferometry probes the internal dynamics of femtosecond soliton molecules," Science **356**, 50-54 (2017).

29. K. Krupa, K. Nithyanandan, U. Andral, P. Tchofo-Dinda, and P. Grelu, "Real-Time Observation of Internal Motion within Ultrafast Dissipative Optical Soliton Molecules," Physical Review Letters **118**, 243901 (2017).

30. Z. Wang, Z. Wang, Y. g. Liu, R. He, G. Wang, G. Yang, and S. Han, "Generation and time jitter of the loose soliton bunch in a passively mode-locked fiber laser," Chinese Optics Letters **15**, 080605 (2017).